# The Vaccine Supply Chain: A Call for Resilience Analytics to Support COVID-19 Vaccine Production and Distribution

Maureen S. Golan

Benjamin D. Trump

Jeffrey C. Cegan

Igor Linkov

US Army Engineer Research and Development Center

Igor.Linkov@usace.army.mil

**Abstract**

Pharmaceutical companies, upstream suppliers, associated logistics providers, health workers, regulatory agencies, public health experts and ultimately the medical practitioners and general public have been navigating an increasingly globalized vaccine supply chain; any disruption to the supply chain may cause cascading failure and have devastating consequences. The COVID-19 pandemic has already highlighted the lack of resilience in supply chains, as global networks fail from disruptions at single nodes and connections. As the race for a COVID-19 vaccine continues, the importance of not only an efficient supply chain but a resilient vaccine supply chain capable of reliable production and reaching target populations despite likely but currently unknown disruptions is imperative. Proactively applying resilience analytics to vaccine supply chain models will increase the probability that vaccination programs meet their goals. Without such a network in place for manufacture and distribution of the COVID-19 vaccine, even the most efficacious and safe vaccine will not prove viable.

Through an overview of the existing vaccine and pharmaceutical supply chain publications focusing on resilience, as well as recent papers reporting modeling of resilience in supply chains across multiple fields, we find that models for supply chain resilience are few and most of them are focused on individual dimensions of resilience rather than on comprehensive strategy necessary for scaling up vaccine production and distribution in emergency settings. We find that COVID-19 resulted in a wave of interest to supply chain resilience, but publications from 2020 are narrow in focus and largely qualitative in nature; evidence-based models and measures are rare. Further, publications often focus exclusively on specific portions of the specific supply chain of interest, excluding associated supporting networks, such as transportation, social and command and control (C2) necessary for vaccine production and equitable distribution. This lack of network analysis is a major gap in the literature that needs to be bridged in order to create methods of real-time analysis and decision tools for the COVID-19 vaccine supply chain. We conclude that a comprehensive, quantitative approach to network resilience that encompasses the supply chain in the context of other social and physical networks is needed in order to address the emerging challenges of a large-scale COVID-19 vaccination program. We further find that the COVID-19 pandemic underscores the necessity of positioning supply chain resilience within a multi-network context and formally incorporating temporal dimensions into analysis through the NAS definition of resilience – plan, absorb, recover, adapt – to ensure essential needs are met across all dimensions of society. Modeling and analyzing vaccine supply chain resilience ensures that value is maintained should disruptions occur.

## I. Introduction

Even in the earliest days of the novel coronavirus SARS-CoV-2 pandemic, considerable public and academic attention has been paid upon the development and distribution of a vaccine that can safely and effectively inoculate the global population. In the absence of other successful preventative measures, public health scholars and policymakers around the world have emphasized this goal is one of achieving "herd immunity," whereby SARS-CoV-2 would not easily be able to spread within a given population given its fundamental biological characteristics and natural reproduction rate. In turn, a critical rate-limiting step to reaching this end goal includes the vast and complex supply chain network underpinning the production and distribution of any vaccine candidate. In this chapter, we review various discourse and challenges related to the robustness and resiliency of pharmaceutical supply chain networks, and note where potential single-points-of-failure may arise that may substantially degrade or even destroy supply chain capacity for successful vaccine development and delivery. In this, we argue that for those points of concern that may jeopardize the sources of any critical vaccine effort, the capacity for system recovery and adaptation must be fostered regardless of the disruption that may be faced in the near or extended future.

As of the beginning of October 2020 there have been more than 34 million cases of COVID-19 (the illness caused by SARS-CoV-2) worldwide with total documented deaths having exceeded one million at the end of September 2020 (WHO 2020b). Complicating the pandemic, co-disruptions such as political unrest and evacuations due to the 2020 storm season decrease the potential to mitigate waves of illness outbreak (Rowland et al. 2020). As cases, death tolls, and socioeconomic consequences rise, finding a viable vaccine is increasingly urgent. As of October 19, 2020, there were 44 candidate vaccines for the novel coronavirus SARS-CoV-2, ranging in approach, from the developer/manufacturer to type of vaccine, number of doses, timing of the doses and route of administration (WHO 2020a). The number of variables in vaccine development add clear complications for planning a large-scale immunization campaign, especially because all but two of the top ten candidate vaccines in Phase III clinical trials as of October 2020 are designed as two-dose vaccines (WHO, 2020b). A two-dose vaccine will increase the burden on the vaccine supply chain, further necessitating the need for a resilient network to be in place and ready to plan, absorb, recover, and adapt to disruptions.

The COVID-19 pandemic highlights the need for building resilient supply chains. In the case of the pandemic, supply chains have been disrupted from all angles. The pandemic caused not a singular disruption, but cascading disruptions affecting multiple nodes, links, and associated networks, including an impact on the demand for certain goods and services. The global nature of supply networks and the movement of goods, information and services has caused unanticipated disruptions for supply chain managers and government agencies. For example, we can look to the global repercussions of China's actions during the onset of the COVID-19 pandemic. The U.S.-China Economic and Security Review Commission found that the stringent restrictions on movement and labor shortages in China caused a sudden drop in China's oil imports, which affected the OPEC supply chain, and caused a disruption to transportation and shipping (Malden and Stephens 2020). This disruption was estimated to affect 75% of U.S. companies, especially impacting electronics, pharmaceuticals, and automotive industry supply chains (Malden and Stephens 2020). An over-reliance on China for electronics manufacturing is a factor cited by DHL's Resilience360 as one of its ten key risks facing global tech supply chains, and ultimately recommend re-evaluating supplier and distribution networks post-COVID-19 (DHL 2020). Similarly, the United States medical and pharmaceutical supply chain is heavily dependent on Chinese manufacturing with a large share of medical and laboratory apparel imported from China (Schwarzenberg and Sutter 2020). Although many of the higher value added products such as biological products and vaccines come from imports (79%), less than 0.05% come from China, with 59% from the European Union under normal circumstances, highlighting the effects that foreign policy as well as dynamics within other countries may have on global supply chains (Elton et al. 2020; Schwarzenberg and Sutter 2020; Sutter et al. 2020).

Echoing this conclusion, a survey of supply chain executives within the first months of the pandemic found that 58% of respondents intend to pivot from single sourcing (Hoek 2020). However, as China's early quarantine lockdowns have slowed the spread of the virus, manufacturing output has increased, but demand and consumption continue to lag, leading to questions concerning inventory and warehousing growth (Feng 2020). This demand uncertainty affects supply chain continuity, as corroborated by an analysis of NASDAQ 100 companies' Twitter feeds spanning from January 2020 to April 2020 (Sharma et al. 2020). As supply chains shift to expand their networks, resilience analytics of the necessary co-networks such as transportation, C2 and Industry 4.0 will need to be evaluated and quantified in tandem with the supply chain itself (Golan et al. 2020).

Geographic sector clustering is also a phenomenon that has supply chain implications. Consequently, the magnitude of a disruption's impact on the domestic supply chain network is traditionally related to the location of origin within the country, as seen in the Coronavirus pandemic (Gomez et al. 2020). Gomez et.al use threshold analysis to find that a disruption in the central U.S. to a supply chain node leads to a larger supply chain failure propagated throughout the country, particularly in the agriculture and food sectors (2020). The effects extend across the border to Canada where border politics and quarantine policies further affect the agriculture supply chains built on just-in-time manufacturing and delivery, and over-land transportation (Hobbs 2020). Understanding the implications of supply chain dependencies on other networks, geographic constraints and policy has been highlighted by the COVID-19 pandemic, supporting our finding for enhanced resilience analytics for vaccine supply chains.

Applying "Industry 4.0" to supply chain management – the move towards more "intelligent" processes, smart technologies, machine learning, digitalization, and overall cyber-physical integration of manufacturing and logistics – allows for big data to be processed in useful manners that can be geared and applied towards resilience and hardening weak points on the supply chain (Golan et al. 2020; Ivanov and Dolgui 2020b; Cavalcante et al. 2019). Collaboration among supply chain tiers and open communication in combination with Industry 4.0 tools such as blockchain, digital supply chain twins, and real-time supply chain updates are coming to the forefront in response to current themes in supply chain resilience (Ivanov and Dolgui 2020b; Sharma et al. 2020; Hobbs 2020; Cavalcante et al. 2019). One especially hard-hit supply chain is the Personal Protective Equipment (PPE), which has also had repercussions on other supply chains due to the inability of people to return to work without proper medical protection. Although gaining mainstream attention during the COVID-19 pandemic, prior analysis foreshadowed this inability for the PPE supply chain to meet demand should a large-scale medical emergency arise (Patel et al. 2017). In the United States, for example, meeting glove, gown, and surgical mask demand will require imports, while the N95 mask, face shield, nasal swab and test kit demand can be (or is expected to be) nearly filled by domestic production, despite the rapidly changing supply chain due to demand increase, non-traditional suppliers, and non-traditional PPE industry users (Elton et al. 2020).

One strategy that is helping to fill this medical supply gap is 3-D printing, which disperses the supply chain. This tool can fill time-critical manufacturing shortages and has been used to source nasal swabs, face shields, and respirator and ventilator components (Salmi et.al. 2020). For example, the National Institutes of Health (NIH) 3D Print Exchange hosts a specific web portal for the *COVID-19 Supply Chain Response*, which includes a collection of PPE designs for clinical and/or community use (NIH 2020). The decentralized aspect of 3D printing, combined with the expansion of accessible technology and Industry 4.0, as well more efficient, and made-to-order aspects of 3D printing show the importance of this technology in supply chains moving forward (Choong et al. 2020). This trend also underscores the notion of private-public cooperation during times of national and global crisis that promote agility in the supply chain, as seen during the 2009 H1N1 and 2014 Ebola epidemic responses (Patel et al. 2017). The PPE supply chain shows the importance of analyzing all network connections in order to understand supply chains because as demonstrated by the COVID-19 pandemic, a failure in necessary supplies of PPE lead to direct and indirect impacts on other supply chains.

As touched upon previously in the geographic perspective, the agriculture and food industries are another example of a sector which has been forced to shift its supply chain due to the COVID-19 pandemic. As with PPE, farming and agriculture supply chains have been disrupted, and the industry has been unable to equitably and adequately meet the basic needs of people across the globe. The COVID-19 pandemic has cascading impacts that are aggravating hunger, hidden-hunger (i.e. malnutrition), and food waste by disrupting access to fresh and affordable foods (Lal 2020). This is an issue of supply chain resilience. Similar to wide-spread use of 3D printing during the pandemic, a more decentralized food system has been suggested to meet current food needs, through trends such as home gardening and urban agriculture (HGUA), delivery apps, and sourcing from local vendors who are less susceptible to border disruptions and labor shortages (Hobbs 2020; Lal 2020). The examples of food and medical supplies in particular highlight the fact that as supply chains are strained, maintaining equitable distribution of essential goods and services must be addressed and ensured through resilience measures.

This strain from the pandemic can be compared to other global stressors. The 2008 global financial crisis has been exemplified to show that stress tests similar to those imposed on U.S. and E.U. banks in the wake of the global crisis can also be advanced by governments to ensure essential supply chains do not fail (Simchi-Levi and Simchi-Levi 2020). The 2008 financial crisis shows how the field of resilience analytics offers a theoretical foundation for policy making in the face of systemic risks and uncertainties (Hynes et.al., 2020b) and can be applied to other complex networks to ensure that critical supplies such as vaccines are available during and immediately after disruptive events. An example of a simple resilience "stress test" is that of Simchi-Levi and Simchi-Levi, which quantifies "time to survive (TSS)" and "time to recover (TRR)," giving a tangible metric for ensuring TSS is greater than TRR (Simchi-Levi and Simchi-Levi 2020). Although overarching regulation might mitigate any cascading failures caused across fragile supply networks, a more complete network model encompassing all "intertwined supply networks (ISN)" is necessary (Ivanov and Dolgui 2020a). This would need to specifically addresses supply chain interdependencies on other networks, as well as trade-offs (e.g. impacts of product cost versus supply continuity) in order to enable true resilience analytics.

The current disruption to existing supply chains due to impacts of the COVID-19 pandemic are evidence of the potential demand for proactive resilience analytics in the COVID-19 vaccine supply chain. A resilient vaccine supply chain will increase the probability of continuous functionality in the face of disruptions, and equitable distribution of the vaccine.

This chapter began by first contextualizing supply chains in general within the systemic threats caused by COVID-19, showing how seemingly unconnected networks have experienced disruptions due to impacts from the pandemic. Next, an overview of the COVID-19 vaccine process is fundamental in understanding the underlying value chain inherent in vaccine production, and the networks that will support resiliency of the system. This section then segues into an overview of the vaccine supply chain and the unique challenges it poses from the manufacturing ecosystem, to the cold chain and the last mile, to reverse logistics and waste management. The existing understanding of the associated networks that enable the vaccine supply chain are also discussed, highlighting the smallpox and MMR vaccination campaigns as case studies, before discussing the existing understanding of disruptions in the vaccine supply chain. Because much of the existing literature and studies on vaccine supply chain is focused on humanitarian response and immunization campaigns, there is a focus on preparing for expected disruptions such as inconsistent power grids in the cold chain. However, unexpected disruptions must also be prepared for as exemplified in the next sections discussing resilience analytics.

The last sections of the chapter transition into a discussion on resilience analytics in supply chain modeling and the clear need for application in the vaccine supply chain. Although the need for modeling resilience in vaccine supply chain is clear, we find a clear lack of focus on this aspect of the vaccine supply chain in existing academic publications, especially in regard to the manufacturing ecosystem. We therefore look at recent trends in modeling supply chain resilience in publications addressing impacts of COVID-19. We then apply these general supply chain modeling trends to the vaccine supply chain and provide recommendations for the COVID-19 vaccine supply chain.

## II. Vaccine Supply Chains

As the global health toll rises, the COVID-19 vaccine is under accelerated development (WHO 2020b; HHS 2020b). Traditional vaccine development proceeds in a linear manner: (1) pre-clinical studies; (2) phase I clinical trials; (3) phase II clinical trials; (4) phase III clinical trials;

(5) infrastructure; (6) manufacturing; (7) approval; (8) distribution and Phase IV post marketing surveillance (WHO 2020b). This linear development allows for reduced risk for stakeholders, infrastructure, and networks associated with the (potential) vaccine as the efficacy and viability of production and clinical studies must be proven before proceeding with each step. However, due to the global nature of the COVID-19 pandemic and the vast toll on human health, the vaccine is under an accelerate development timeline, which overlaps the steps in vaccine development: (1) pre-clinical, phase I, phase II, phase III, infrastructure, manufacturing; (2)

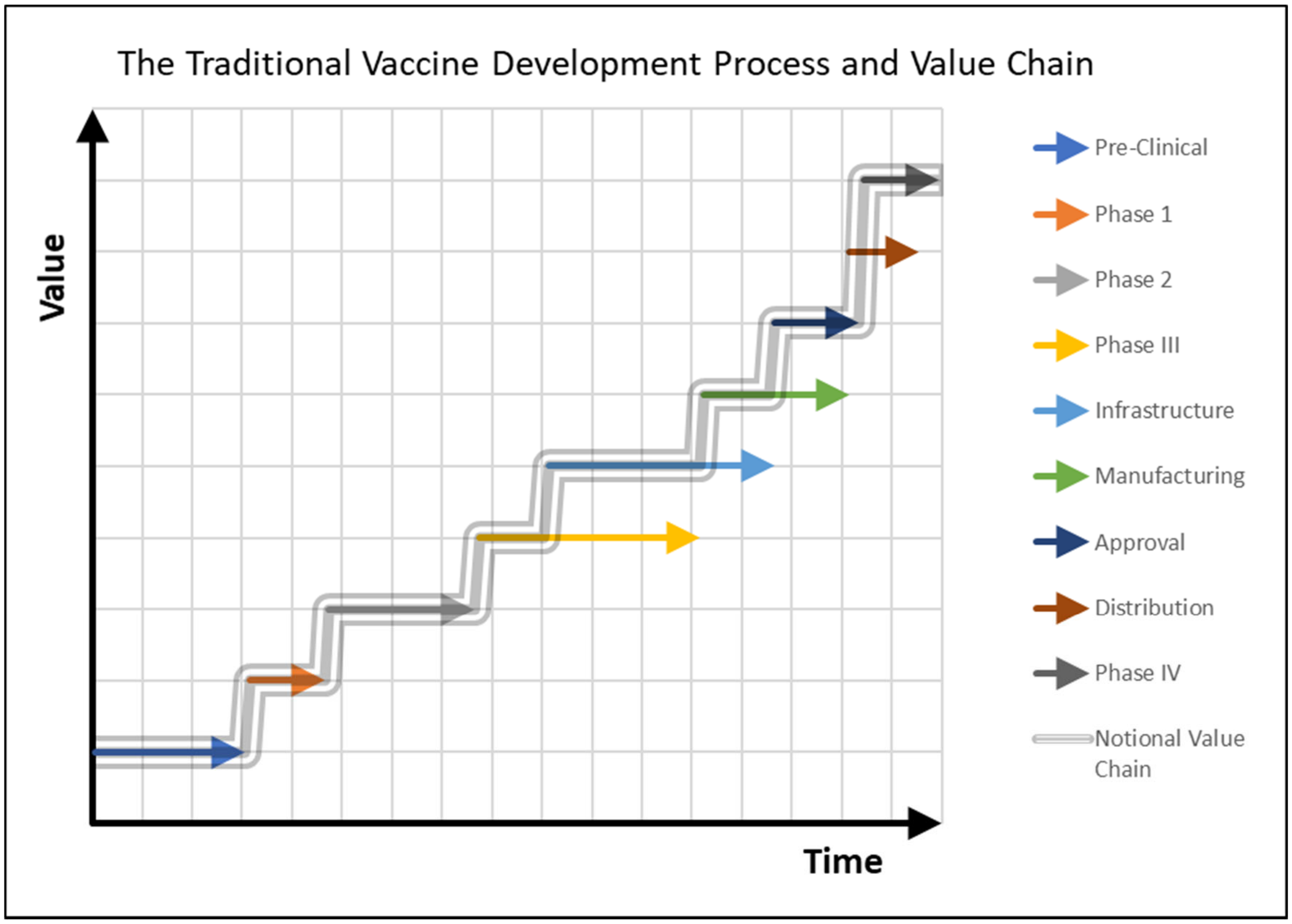

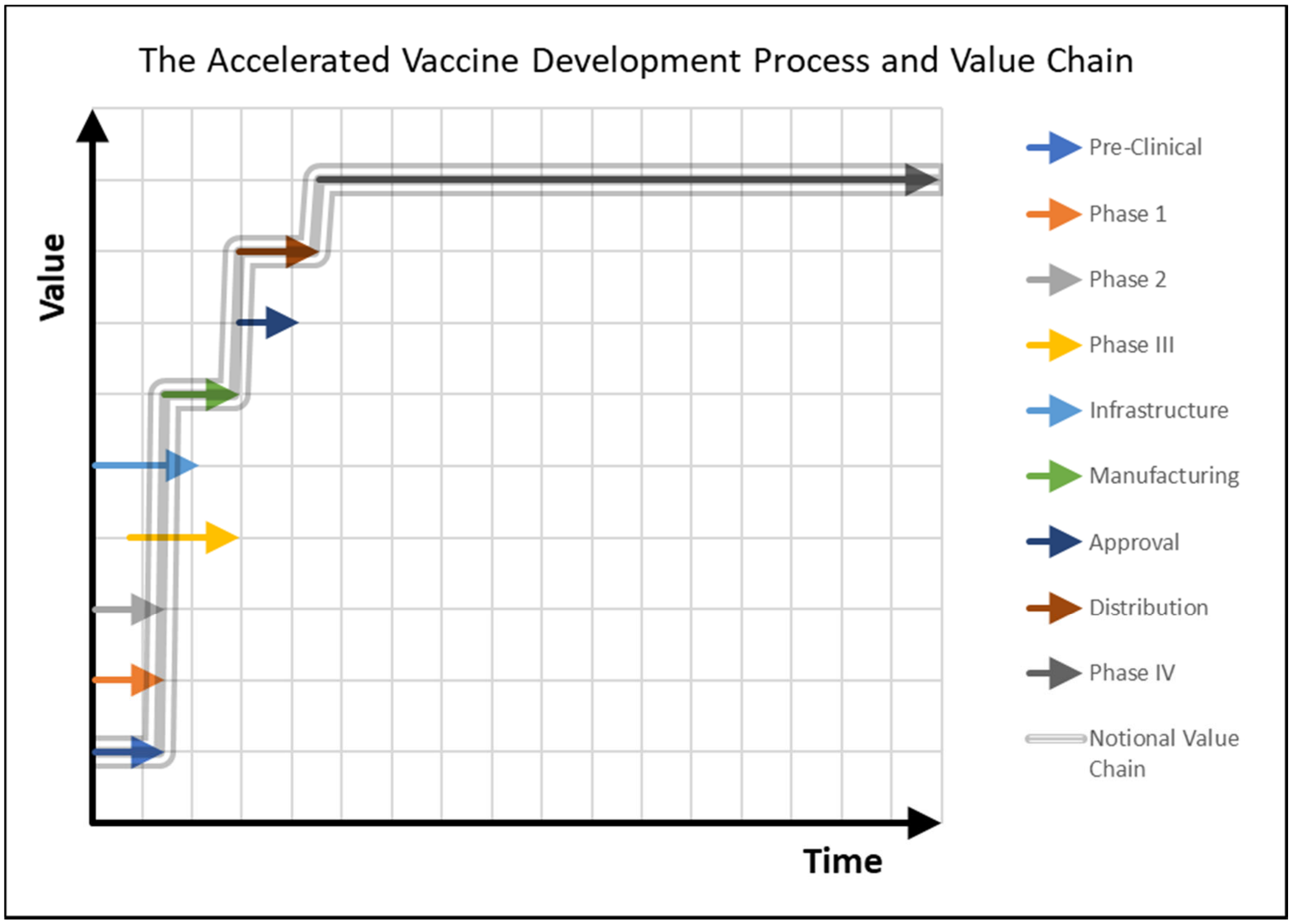

*Figure 1: The traditional (a) and accelerated (b) vaccine development processes in relation to the value chain (see Linkov et. al. 2020 for discussion of value chain and resilience).*

approval, distribution; (3) phase IV (WHO 2020b). Applying the notion of value chain, whereby the "value" of a product increases exclusive of initial costs in order to gain competitive advantage (Porter 1985), the accelerated vaccine development (*Figure 1b*) gains value much faster than the traditional vaccine development (*Figure 1a*). In both cases, in order for a successful vaccine deployment to occur, and to maintain continued optimal performance, resiliency must be built into the system.

This unprecedented acceleration and volume of vaccine required furthers the burden on the supply chain. For this reason, many of the first proposed treatments and vaccine studies looked at existing medicines, such as ribavirin, due to existing inventories and reliable supply chains (Khalili et al. 2020). However, as novel techniques for vaccine manufacture are investigated and implemented in an accelerated process, unprecedented strain and risk will be placed on manufacturing (Graham 2020). Vaccine manufacturing is already more challenging than typical small molecule pharmaceuticals due to the compound risk of physical and biological variability, different populations approved for each vaccine (e.g. pregnant women, infants, immunocompromised persons), level of antibody response, and side effects, leaving many vaccine manufacturers to fail despite the unmet immunization demand (WHO 2020b; Plotkin et al. 2017). Although understood to be a risky business despite protocols enumerated by regulatory and governing agencies, insight into infrastructure and distribution networks beyond the manufacturing process itself – the network of underlying supply chains – that will ultimately be required for the future vaccine must be analyzed. A resilient vaccine supply chain must be implemented to ensure that once a vaccine is ready for Phase IV, populations will be able to access the vaccine in an equitable manner.

The viability and efficacy of a COVID-19 vaccine must be analyzed in tandem with the networks that will create value and support network resilience post disruption. Without an efficient supply chain capable of withstanding disruptions (i.e. resilient), even the most effective vaccine will be rendered ineffective at disease prevention.

*2.1. Vaccine Supply Chains*

In general, pharmaceutical manufacturers depict vaccine production semi-linearly from raw material reception, active ingredient manufacturing, coupling and formulation, filling, packaging and lot release, shipment, and distribution (Sanofi 2019; VE 2016). In broader terms, global immunization programs, such as the GAVI Alliance, which promotes increased access to vaccines in developing countries, underscores the importance of supply chain through promotion of its five pillars: people and practice, cold chain equipment, data for management, distribution and system design (Brownlow and Thornton 2014). In sum, the efficacy of the vaccine depends not only on the supply chain that distributes the final product, but the supply chains underpinning the manufacture itself. The importance of the manufacturing ecosystem is highlighted by the fact that both the specific biological entity and the specific production process are regulated, with even subtle changes to consumables compromising final safety, efficacy and purity, meaning that even if the product remains theoretically identical, if the process changes, clinical trials and licensing will need to occur again (Plotkin et al. 2017). This is highlighted by the fact that many vaccine patents protect the manufacturing process rather than the active antigen produced by that process (Plotkin et al. 2017). Therefore, a steady supply chain of constant and high-quality materials (i.e. consumables) is crucial from the beginning of the vaccine supply chain.

Understanding the entire vaccine supply chain leading to manufacture and from manufacture to the last mile is essential to modeling the entire network, with all its interactions. Therefore, this section, breaks out the vaccine supply chain into *manufacturing*, which addresses the upstream and downstream consumables through secondary packaging (Section 2.1); *cold chain*, which addresses the distribution challenges of vaccine supply chains (Section 2.2); *last mile*, which addresses the fundamental goal of the vaccine supply chain: meeting immunization targets (Section 2.3); *reverse logistics and (biohazard) waste management*, which addresses the additional supply chains nodes and links necessary for disposal of medical waste and returned cold chain equipment (Section 2.4); and *vaccine supply chains and interconnected networks*, which addresses the network qualities of the vaccine supply chain and its dependence on associated networks such as transportation and command and control (C2) (Section 2.5).

Similar to the vaccine overview provided, this section is intended as a basis for understanding vaccine supply chain complexities and constraints in relation to the current state of vaccine supply chain models, and areas that need to be addressed for a large-scale pandemic inoculation campaign to maintain optimal performance. Although broken out into sub-sections, there is inherent overlap, and is meant as a tool for highlighting components of the vaccine supply chain.

### *2.1 Manufacturing*

The average vaccine manufacturing process takes from 6-36 months due to not only the complexity of the biopharmaceuticals themselves, but also the quality control processes, which represent about 70% of the manufacturing duration (Sanofi 2019; VE 2016). The production of each vaccine is unique, but generally includes (1) raw material reception, with some vaccines requiring ~160 unique consumables; (2) bacteria, virus, or cell culture; (3) harvesting; (4) purification; (5) inactivation; (6) valence assembly; (7) formulation; (8) filling; (9) freeze-drying; (10) packaging; (11) batch release (Sanofi 2019; VE 2016). These steps in turn rely on their own raw materials supply chains, such as glassware and pharmaceuticals, as well as platform specific factory capabilities (Rele 2020). Although these processing stages are ultimately monitored for efficacy and human health guidelines through strict licensing requirements, the mechanics of the manufacturing supply chains, from the complexity, challenges and costs, is not readily available in the existing literature (Plotkin et al. 2017). This is partly due to the inherently challenging nature of vaccine manufacture, as outcome variability can be high from "the nearly infinite combinations of biological variability in basic starting materials" with factors ranging from the microorganism, the environmental condition of the microbial culture, the experience and knowledge of the manufacturing technician, the steps involved in the purification process, and even the analytical methods used to measure the biological processes and antigens present (Plotkin et al. 2017).

The Bill and Melinda Gates Foundation compartmentalizes the manufacture of commercial vaccines into four specific economic steps which are useful in consideration of the supply chain: (1) bulk: downstream and upstream production costs; (2) form/fill/finish: final dosage form and formulation (adjuvantation, lyophilization, etc.), aseptic filling, and remaining production finishing steps (vial labeling, etc.) costs; (3) secondary packaging: carton distribution preparation and warehousing costs; (4) QC/QA: quality control and quality assurance testing costs, which may occur at multiple points in manufacturing (Iqbal and Lorenson 2016). Within the bulk production, the upstream process is whereby the immunogen, the antigen that elicits an immune response, is generated through a cultivation process, and the downstream process is the purification process whereby the host cell impurities are removed and additional processing occurs to result in a bulk vaccine (Pujar et al. 2014). Overall quality controls ensure that the vaccine conforms to release specifications throughout its entire life cycle, including long-term stability tests required on several lots each year, adherence to all applicable National Regulatory

Authorities (NRAs) and Current Good Manufacturing Processes (cGMP), maintaining proper process documentation and ratios of Quality Assurance personnel to production personnel (Plotkin et al. 2017).

Underlying each step of the manufacturing process is the consumable, the raw materials used as inputs in vaccine production, and consequently, their sourcing play a major role in the manufacturing supply chain and economics of vaccine development and production (Iqbal and Lorenson 2016). These consumables, stemming from biological processes themselves, are subject to variability in manufacturing, processing or contamination (i.e. materials of animal origin carry the risk of adventitious agents – viruses or other microbes – and therefore often sourced from disease-free regions) (Plotkin et al. 2017). Items such as facilities, equipment, direct labor and overhead including indirect overhead –plant management salaries, wages, training – and corporate overhead – executive salaries, centralized back-office functions, insurance – must also be accounted for as components or variables affecting the vaccine manufacturing supply chain, as well as commercialization and licensing (Iqbal and Lorenson 2016). Despite equipment commonalities across vaccine platforms, the specific cycles and sequences of operations varies to the extent that each product usually has a dedicated facility and group of experts, which allows for flexibility with unpredictable demand at the trade-off of higher cost (Plotkin et al. 2017). Furthermore, there is a global scarcity of personnel and labor with the skills, knowledge and expertise required by the vaccine industry, especially Quality Assurance personnel (Plotkin et al. 2017). Labor is also a necessary component of the supply chain that cannot be overlooked and must be factored as a component of resilient vaccine manufacture and distribution networks.

*Table 1* provides an overview of the manufacturing supply chain and the general ecosystem surrounding vaccine production supply chains. It also shows constraints or stressors that may lead to disruptions in the manufacturing supply chain. Every vaccine is developed uniquely and as such this table is neither meant to be all-encompassing nor represent every vaccine, but to highlight general patterns in the supply chain as well as the inherent challenges in manufacture and scaling a vaccine supply chain. Platform vaccines do offer unique capabilities for efficient vaccine development, but due to their novelty, a large-scale manufacturing process has not yet been proven world-wide (HHS 2020d; Plotkin et al. 2017; Pujar et al. 2014). Consequently, strains on the supply chain have not been tested and hardened again risk and disruption.

Vital to vaccine manufacture is the cell line or culture that will be used. Historically, this has been conducted *in vivo* and *in ovo* meaning that vaccines still using these processes rely on sufficient supplies of live and pathogen-free eggs and animals, which can only accomplish scale-up through scale-out (i.e. increasing number of eggs) while using automation (Pujar et al. 2014). The *in vitro* production breakthrough, which came with the Polio vaccine, has been significant for industrial production that can now utilize either primary cells or continuous cell lines (e.g. Vero from monkey kidney cells, chick embryo fibroblasts, WI-38 and MRC-5 human diploid cells, MDCK from Mardin-Darby canine kidney) and cell-culture-based production systems, allowing for scale-up and relatively rapid shifts in production (Pujar et al. 2014). Furthermore, cell banking became a key feature of biomanufacturing allowing for stable and well-tested substrates for each vaccine batch, but require adherent surfaces for growth, which similarly limits scale-up with scale-out of surface area (Pujar et al. 2014). As such, microcarriers rather than bioreactors for production of new rabies and polio vaccines have been used for industrial purposes, with new vaccines now able to use serum-free media and suspension culture that is more conducive to scalability (Pujar et al. 2014). Each of these shifts in the upstream processing have simultaneously necessitated shifts in the underlying supply chains and changing biopharmaceutical technology. Likewise, corresponding changes to the downstream processing have also occurred (Pujar et al. 2014).

*Table 1: General Vaccine Production and Ecosystem Surrounding Vaccine Manufacturing Supply Chains (Sanofi 2019; Plotkin et al. 2017; Iqbal and Lorenson 2016; VE 2016; Pujar et al. 2014).*

| | Bulk | Form/Fill/Finish | Packaging & Lot Release | QA/QC |
|---|---|---|---|---|
| **Material** | • Biological agents (e.g. yeast extract, E. coli, natural / recombinant enzymes)<br>• Chemical agents<br>• Appropriate continuous cell lines & substrate<br>• Associated raw materials & consumables for vaccine<br>• Bioreactor (roller bottles, T-flasks, cell cubes, cell factories)<br>• Microcarriers & serums<br>• Centrifuges<br>• Column & absorbent (e.g. alumina)<br>• Sterilization materials | • Vials<br>• Syringes<br>• Needles<br>• Stoppers<br>• Seals<br>• Buffers<br>• Adjuvant<br>• Stabilizers<br>• Membrane filters<br>• Preservatives (e.g. formalin)<br>• Associated raw materials & consumables for drug product | • Labels (country specific)<br>• Leaflets (country specific)<br>• Vaccine Vial Monitors (VVM)<br>• Secondary and tertiary cartons | • Inputs for testing kits<br>• Assays<br>• Documentation system of cGMP |
| **Operational** | • Culture, harvest & extraction<br>• Fermentation<br>• Purification<br>• Inactivation & valence assembly<br>• Centrifugation<br>• Chromatography<br>• Contamination controls<br>• Direct & indirect labor<br>• IT systems<br>• Facilities & maintenance<br>• Raw material delivery (transportation & infrastructure)<br>• Security<br>• Material sourcing (logistics)<br>• Inventory management<br>• Utilities | • Filtration<br>• Stabilization (e.g. pH management, lyophilization)<br>• Adjuvantation<br>• Aseptic filling & crimping<br>• Contamination controls<br>• Direct & indirect labor<br>• IT systems<br>• Facilities & maintenance<br>• Raw material delivery (transportation & infrastructure)<br>• Security<br>• Material sourcing (logistics)<br>• Inventory management<br>• Utilities | • Single-product vs. multi-product operation<br>• Direct & indirect labor<br>• IT systems<br>• Facilities & maintenance<br>• Raw material delivery (transportation & infrastructure)<br>• Licensing<br>• Security<br>• Material sourcing<br>• Inventory & distribution management<br>• Utilities | • Enhanced process control (e.g. PAT, QbD)<br>• IT systems<br>• Raw material delivery (transportation & infrastructure)<br>• Security<br>• Material sourcing<br>• Inventory management<br>• Utilities<br>• Quality Assurance Personnel<br>• Testing by manufacturer (raw material reception, API production, coupling & formulation, filling, packaging)<br>• Testing by exporting country (API production)<br>• Testing by importing country (packaging & lot release) |
| **Constraint(s)** | • Shelf-life<br>• Biological variability<br>• Processing scale<br>• Facilities & equipment<br>• Manufacturing dynamics & supply of consumables<br>• Batch failure<br>• Consumable quality & reliability<br>• Tariffs / Nationalism / closed borders due to quarantine | • Shelf-life<br>• Biological variability<br>• Processing scale<br>• Facilities & equipment<br>• Consumable quality & reliability<br>• Batch failure<br>• Consumable quality & reliability<br>• Tariffs / Nationalism / closed borders due to quarantine | • Shelf-life<br>• Storage / warehousing<br>• CTC or cold chain requirements<br>• Facilities & equipment<br>• Batch failure<br>• Country-specific & population-specific labelling requirements<br>• Tariffs / Nationalism / closed borders due to quarantine | • Variability within QA/QC testing<br>• Licensing<br>• Regulatory requirements<br>• Inexperienced labor / regulators (with specific vaccine, product, regulation, etc.)<br>• Tariffs / Nationalism / closed borders due to quarantine<br>• NRA enforcement /compliance change / country specific import / clinical trial requirements |

However, the shifting technology available to the vaccine industry, does not mean that "historic" vaccines have stopped being produced, having impacts on the supply chain as well. Most vaccines in use today were developed in the 1940s and 1950s and underscores that necessity of stable raw quality materials and component supplies from reliable vendors (Plotkin et al. 2017; Pujar et al. 2014). In addition to the cumbersome regulatory process, moving pharmaceutical manufacturing offshore has also heightened the barrier to innovation, putting competitive advantage in cheap labor and environmental pollution regulations rather than new technology (Gurvich and Hussain 2020; Plotkin et al. 2017). These existing vaccine supply chains must be able to operate concurrently with ramping of new vaccines for pandemics, or other disruptions to the existing supply chains underpinning vaccine production.

In order to provide greater specificity on the supply chain, the vaccine type must be considered: live-attenuated vaccines, inactivated vaccines, live or whole-killed bacterial vaccines, subunit, recombinant, polysaccharide, and conjugate vaccines, toxoid and classical subunit vaccines, DNA vaccines, and recombinant vector and platform based vaccines (HHS 2020d; Pujar et al. 2014). Vaccines may also be multivalent, increasing their manufacturing complexity (Pujar et al. 2014) and as a result, increasing supply chain complexities. *Table 2* further breaks down these vaccine specifics within the manufacturing supply chain ecosystem.

*Table 2: Standard Vaccine Classifications and General Manufacturing within the Overall Vaccine Supply Chain Tiers / Network. Note that this table is meant as an overview to highlight overarching supply chain dynamics; every vaccine manufacturing process is uniquely developed and subsequently approved by NRAs for specific populations and must be identified individually for greater nuance in the manufacturing ecosystem (Folegatti et al. 2020; Graham 2020; HHS 2020d; van Riel and de Wit 2020; Adalia et al. 2019; Pujar et al. 2014).*

| Vaccine Type | Examples | Notable Characteristics Affecting Vaccine Supply Chain | Upstream | Downstream |
|---|---|---|---|---|
| Live-attenuated | • Measles, mumps, rubella<br>• Rotavirus<br>• Smallpox<br>• Chickenpox<br>• Yellow fever | • Uses weakened form of virus – risk to immunocompromised<br>• Complex upstream cultivation & minimal downstream processing<br>• Scalability through continuous cell lines, microcarriers and serum-free suspensions<br>• Life-time immunity<br>• Cold chain | • Host / cell substrate choice vital – impacts to vaccine safety & reactogenicity<br>• *In ovo* production still in use for some<br>• Continuous cell lines / banks monitored for oncogenicity | • Harvest technologies & column chromatography for high levels of purification<br>• Enhanced stabilization through lyophilization (but also risk of loss of potency)<br>• Continued focus on lyophilization for improved thermostability |
| Inactivated | • Hepatitis A<br>• Influenza (flu)<br>• Polio<br>• Rabies<br>• Japanese encephalitis | • Uses dead form of the virus<br>• Booster shots required<br>• Scalability through continuous cell lines, microcarriers and serum-free suspensions<br>• Cold chain | • Host / cell substrate choice vital – impacts to vaccine safety & reactogenicity<br>• *In vivo* production still in use for some<br>• *In ovo* production still in use for some<br>• Continuous cell lines / banks monitored for oncogenicity<br>• Microcarriers for continuous cell line scale-out over bioreactors<br>• Serum-free media and suspension for scalability<br>• Development of "immortal" recombinant (designer) cell lines | • Harvest technologies & column chromatography for high levels of purification<br>• Enhanced stabilization through lyophilization (but also risk of loss of potency)<br>• Continued focus on lyophilization for improved thermostability<br>• Most require adjuvants (e.g. aluminum salts) for enhanced immunogenicity, but increase manufacturing complexity |
| Live or whole-killed bacterial | • Cholera<br>• Typhoid fever<br>• First whooping cough vaccine | • Simplest manufacturing bioprocessing of all vaccines<br>• Lyophilization advancement for typhoid vaccine solid dosage form<br>• Scalability through stirred tank fermenters with process monitoring & controls<br>• Cold chain | • Cultivation of bacteria and harvesting<br>• Inactivation as necessary with heat<br>• Fed-batch fermentation for scalability – chemically defined medium in stirred tank fermenters (with aeration for aerobic cultures) | • Control of toxin production during fermentation<br>• Suspensions formulated with formaldehyde<br>• Enhanced process control through process analytical technology (PAT), Quality by Design (QbD)<br>• Enhanced stabilization through lyophilization (but also risk of loss of potency)<br>• Continued focus on lyophilization for improved thermostability |
| Subunit, recombinant, polysaccharide, conjugate | • Hepatitis B<br>• Hib (Haemophilus influenzae type b) disease<br>• HPV<br>• Whooping cough (DTaP combined vaccine)<br>• Pneumococcal disease<br>• Meningococcal disease<br>• Shingles<br>• Anthrax | • Uses specific pieces of virus – low risk to immunocompromised, but lower inherent immunogenicity<br>• Acellular subunit vaccines address reactogenicity of the whole vaccine<br>• Scalability through stirred tank fermenters with process monitoring & controls<br>• Pnumovax®23 developed in the 1980s is the broadest valency vaccine<br>• Cold chain | • Early media containing animal components (e.g. beef digest medium)<br>• Shift to dairy media (e.g. casein digest), and to complete animal-derived component removal<br>• Recombinant subunit requires rDNA technology<br>• Conjugate reaction dependent on two high value intermediates; minimization of side reaction vital; scale-up through chemical processing (e.g. control rates of reaction via fluid transport) but complexity of vaccines ultimately limits | • Removal of cellular components and process residuals (extent varies with vaccine)<br>• Purification may require precipitation using CTAB<br>• High likelihood of contamination / intrusion requires batch fermentation & special facilities to prevent biocontainment<br>• Most require adjuvants (e.g. aluminum salts) for enhanced immunogenicity, but increase manufacturing complexity |
| Toxoid (classical subunit) | • Diphtheria<br>• Tetanus | • Uses virus toxins<br>• Booster shots required<br>• Scalability through stirred tank fermenters with process monitoring & controls<br>• Scalability through toxoid purification from precipitation and low-resolution chromatography to tangential flow membranes<br>• Cold chain | • Early media containing animal components (e.g. beef digest medium)<br>• Shift to dairy media (e.g. casein digest), and to complete animal-derived component removal<br>• Fermentation, harvest, centrifugation, and inactivation with formalin | • Extensive protein purification – ammonium sulfate precipitation, chromatography, membrane filtration<br>• Most require adjuvants (e.g. aluminum salts) for enhanced immunogenicity, but increase manufacturing complexity |
| Recombinant vector & DNA (platform based) | None to date | • Theoretically adaptable, streamlined and inexpensive with the same downstream / upstream processes for each platform (economies of scale not yet proven)<br>• Long-term immunity<br>• Scalable<br>• NRAs continue to license product not platform<br>• DNA poses challenging delivery methods<br>• Cold chain (possible exception of DNA) | • Underlying identical mechanism, device, cell line, or delivery vector for multiple target vaccines (e.g. chimpanzee adenovirus)<br>• Spectrum of platforms (VSV, ChAd, 17D, MVA, Baculovirus expression system, Tobacco plant cells, synthetic mRNA, DNA, self-amplifying RNA, nucleic acid printer)<br>• mRNA-based provides additional manufacturing and biological delivery capabilities; target adaptability | • Purification required accounting for spectrum of platform types – viral vectors, expression platforms, nucleic acids, nucleic acid printer<br>• Continuous antigen production limits to prevent desensitization |

Overall, there are very few academic or publicly available industry publications focusing on the specifics of the manufacturing supply chain, and those that do generally focus on aspects pertaining to either economics of manufacturing, the global humanitarian immunization effort, or a combination of both. Publications looking at modeling the manufacturing supply chain are even more limited. For example, a Web of Science "All Databases" topic search with no time limit (i.e. 1864 to November 1, 2020) and no citation requirements for "vaccin*" AND "manufactur*" AND "supply chain*" AND "model*" resulted in 28 publications. Of these, 20 were relevant to vaccine supply chain manufacturing and focused on modeling aspects of the network. These publications are shown in *Table 3*. None of these publications specifically analyze resilience, although one focuses on risk by limiting the intricacy of the supply chain modeled and employing entropy modeling to increase public benefit to influenza campaigns (Levnar et al. 2014). Although many of the publications provide quantitative analysis and develop supply chain models, their focus remains limited to vaccine types, countries, specific portions of the manufacturing (or overall) vaccine supply chain, meeting immunization targets without actual analysis of the manufacturing supply chain, or non-inclusive of associated networks. In order to harden vaccine supply chains against disruptions, they first need to be modeled using network analytics, starting with manufacturing, and then understood through resilience analytics. (Note that there are two publications, however, under the larger topic vaccine supply chain resilience, that are returned when eliminating "model*" and "manufactur*" and this is discussed in Section 3.1.)

*Table 3: WOS Relevant Publications with Vaccine Manufacturing Supply Chain Models.*

| Publication | Vaccine Focus | Global Focus | Process Focus | Market Focus | Quantitative / Qualitative | SC Model | Associate Network Representation or Discussion |
|---|---|---|---|---|---|---|---|
| Martin, P. et al (2020) | Late-stage vaccines | Developing Countries | Distribution | Contract – Global Healthcare Organizations | Qualitative | None – economic discount pricing | None |
| Graham, B. S. (2020) | COVID-19 | World-wide | Research & Development | Global Pandemic | Qualitative | None | None |
| Dixit, A. et al. (2019) | N/A – Literature Review | N/A | Distribution, cold chain (manufacture excluded) | Healthcare supply chains | N/A | N/A | None |
| Souza, L. et al. (2019) | National Immunization Program | Brazil | Distribution, cold chain | Logistics | Qualitative | None | Transportation |
| Wedlock, P. et al. (2018) | Measles – rubella (MR) | Benin, India, Mozambique | Cold chain storage & wastage (reverse logistics) | Immunization campaigns | Quantitative – trade-offs between cost per dose & administered amount | HERMES-generated immunization SC model | Other vaccine networks (i.e. cold chain bottlenecks) |
| Ouzayd, F. et al. (2018) | National immunization program | World-wide | Cold chain | Real-time tracking | Quantitative | SCOR model & Colored Petri Net Theory | Transportation |
| Hovav, S. and Herbon, A. (2017) | Large-scale influenza vaccination program | World-wide; Israeli Case Study | Distribution | Healthcare Organization / Services | Quantitative – trade-offs with cost & population coverage | Mixed-integer programming optimization model | Transportation |
| Chick, S. et al. (2017) | Influenza vaccination program | Conceptual | Procurement | Contract – Government Healthcare Programs | Quantitative – trade-offs with production yield, procurement, government oversight resources | Game theory – optimization with asymmetrical information | None |
| Thompson, K. and Tebbens, R. (2016) | Global immunization programs; measles and cholera case studies | World-wide | Universal vaccine stockpiles | Policy - Global Healthcare Organizations | Qualitative | Stock-and-flow | Transportation; Other vaccine interventions |
| Hansen, K. and Grunow, M. (2015) | Secondary pharmaceutical production (i.e. downstream of API) | Conceptual | Manufacturing Ramp-up Capacity | Pharmaceutical Industry | Quantitative | Mixed-integer linear programming | Construction time |
| Chen, S. et al. (2014) | Global immunization programs | Developing countries; Niger, Thailand, Vietnam case studies | Distribution networks | WHO's Expanded Program on Immunization (EPI) | Quantitative | Linear programming | Transportation; capacity expansion |
| Levner, E. et al. (2014) | Large-scale influenza vaccination program | World-wide; Israeli Case Study | Supply Chain Risk | Healthcare Organization / Services | Quantitative – trade-offs with manufacturing, distribution, inventory costs with public benefit | Integer programming on reduced SC with extension of classical Shannon's entropy concept (information theory); GAMS software | Transportation |
| Anderson, R. et al. (2014) | Global immunization programs | Laos | Cold Chain Information System (CCIS) | UNICEF and Lao Ministry of Health | Quantitative – real time measurements of inventory & temperature | None | None |
| Norman, B. et al. (2013) | Global immunization programs (set list of vaccines) | Conceptual network | Cold chain – passive cold storage devices (PCDs) | WHO's Expanded Program on Immunization (EPI) | Quantitative – economic trade-offs between portable/stationary PCDs, and solar refrigerators | Graph SC with economic scenarios | Transportation |
| Arifoğlu, K. et al. (2012) | Influenza vaccination | Conceptual networks | Allocation of Final Product | Social planners (e.g. governments) | Quantitative – comparative economic interventions | Numerical equilibrium analysis | None |
| Lee, B. et al. (2012) | Pneumococcal (PCV-7) and rotavirus vaccines (RV) | Niger | Routine immunization programs (i.e. impacts on other vaccine SCs) | WHO's Expanded Program on Immunization (EPI); Bill and Melinda Gates Foundation | Quantitative – supply ratios | Linear SC model | Transportation; Other vaccine networks |
| Lee, B. et al. (2011a) | Pneumococcal (PCV-7) and rotavirus vaccines (RV) | Thailand | Routine immunization programs (i.e. impacts on other vaccine SCs) | WHO's Expanded Program on Immunization (EPI); Bill and Melinda Gates Foundation | Quantitative – baseline vaccine comparisons | HERMES-generated SC & deterministic mathematical equation-based modeling | Transportation; Other vaccine networks |
| Smith, G. et al. (2011) | Measles containing vaccine (MCV) | World-wide | Manufacturing capacity | World Health Organization | Quantitative – population dynamics & projected supply | None | None |
| Assi, T.-M., et al. (2011) | Measles | Niger | Distribution – vial size | WHO's Expanded Program on Immunization (EPI); Bill and Melinda Gates Foundation | Quantitative – trade-offs with cost, storage, availability, transport, wastage, disposal | HERMES-generated SC, discrete event simulation model | Transportation |
| Lee. B., et al. (2011b) | Measles (MMR) | Thailand | Distribution – vial size | WHO's Expanded Program on Immunization (EPI); Bill and Melinda Gates Foundation | Quantitative – trade-offs with wastage, administration costs, and reserve decrease | Linear supply chain | Transportation |

One strategy to improve the resilience of the vaccine supply chain is for modular manufacture. For example, the Bill and Melinda Gates Foundation focuses on strategies for both novel delivery formats as well as modular, automated manufacturing platforms enabling small-batch vaccine production (Gates 2020). However, the benefit of large-scale manufacturing of vital vaccines has been revolutionary for global immunization targets (Pujar et al. 2014). Conversely, the high cost of entry for vaccine manufacture limits the potential profit of developing countries to invest and produce their own vaccines due to the equipment, personnel, consumables necessary (Plotkin et al. 2017). Although this would be a similar shift in the supply chain as that seen in 3D printing for PPE during supply chain disruption, the vaccine industry is significantly regulated and vaccine safety and efficacy of paramount importance.

Another strategy that can be employed at the manufacturing stage of vaccine supply chains to help buffer against sourcing disruptions is stockpiling. However, this has been more commonly used as a tool for finished products, where distribution scenarios (last mile) may impact the progression of pandemic spread (Davey et al. 2008). Long-term focus on commercialization optimization within the manufacturing of vaccines and preceding tiers rather than manufacturing optimization (i.e. consideration of all the networks) can similarly decrease risk of disruption, prioritizing long-term cost savings over short-term revenue (Plotkin et al. 2017). This includes ensuring that stable manufacturing supply chains are available for commercialization so that immunization goals are met. Understanding the entire supply chain through network analytics is imperative to long-term success. Given the risks in manufacturing and unanticipated (or even anticipated) disruptions, a more resilient manufacturing supply chain will be able to continue converging its operational maximum capacity to its theoretical maximum capacity post disruption, causing less disruptions in the remainder of the supply chain.

*2.2 Cold Chain*

As most vaccines must be kept within narrow temperature ranges between 2°C and 8°C, a specific "cold chain" process for manufacturing, distribution, storage and administration of these vaccines to ensure potency, effectiveness and ultimate safety for populations is necessary, beginning with temperature control at the manufacturing plant and extending through transportation and administration to the patient (Kumru et al. 2014; CDC 2019; UN 2020). The CDC further breaks the cold chain into responsible agents, with the manufacturer responsible for temperature control at the "vaccine manufacturing" stage, the manufacturer and distributor both responsible at the "vaccine distribution" stage, and the provider responsible at the "vaccine arrival at provider facility," "vaccine storage and handling at provider facility," and "vaccine administration" stages (CDC 2019). In other words, any disruptions in the cold chain are considered off-label use, putting liability on the individual practitioner (Purssell 2015). Even if the equipment for the cold chain is in place, proper protocols and procedures must be in place among the adjoining networks – personnel must be trained properly and transportation available.

Having a cold chain that is capable of responding to different vaccine requirements is essential, as some vaccines such as live, attenuated viral vaccines are sensitive to elevated temperatures, whereas some cannot be frozen such as aluminum adjuvanted vaccines (Kumru et al. 2014). The cold chain aims to limit vaccine exposure to inappropriate conditions due to irreversible impacts to potency if outside recommended temperature ranges, as well as complete destruction of potency if frozen (CDC 2019). There is often no visible evidence that potency has been lost or destroyed and can only be determined through expensive laboratory assays only cost-effective at scale (Galazka et al. 1998; CDC 2019). Although the cold chain poses elevated challenges in warmer climates with longer distribution legs, the challenge lies in keeping the product within a constant temperature range, not necessarily the cold temperature itself (Gunn 2020).

Most pharmaceutical companies rely on service providers for distribution of temperature dependent products. Pfizer, for example, coordinates logistics through a system of rigorous oversight and audits with third-party distributors (Gunn 2020). However, aid organizations, such as Doctors Without Borders (MSF) have set up an end-to-end supply chain and coordinate the entire vaccine distribution process (Gunn 2020). The MSF biopharmaceutical supply chain is based on three supply centers in Europe, before passing through the technically challenging last mile of the cold chain that could be thousands of miles long (Gunn 2020). In the case of the Ebola epidemic, the rVSV-ZEBOV vaccine by Merck has a 97.5% effectiveness rate, but must be stored at -70°C to -80°C, leaving logistics and supply chain implementors to use ice-lined generator- and solar-powered refrigerators that can meet temperature requirements during power failure (Gunn 2020).

The MSF Ebola vaccine distribution highlights the use of both passive and active cold chain equipment (CCE) used for storing and transporting vaccines. Active CCE includes mains refrigerators, which are cooled through compressors powered by an existing electric grid, and off-grid refrigerators, which are cooled through either absorption (burning liquid petroleum gas or kerosene) or solar powered compressors (solar battery-powered or solar direct-drive) (Robertson et al. 2017; Chen et al. 2015). In passive CCE, the cooling is provided through coolant packs of phase changing material (PCM) and includes devices such as cold boxes and vaccine carriers (Robertson et al. 2017; Chen et al. 2015). CCE is so vital to the vaccine supply chain, that the World Health Organization has a list of approved CCE and recommended uses available to the public (WHO 2012). These approved devices are not a cure-all, however, and must be used properly. The passive cooling devices, for example, will cause freeze damage to vaccines if conditioning of the coolant packs is not conducted prior to placement of the vaccines (Robertson et al. 2017).

Emerging technologies such as remote temperature control and satellite tracking are already being used to manage risk in cold chain operations (Ouzayd et al. 2018; Gunn 2017; Anderson et al. 2014). Incorporation of remote sensors in vaccine shipments allows for targeted intervention and better understanding disruptions to the cold chain. One company, Parsyl, found that the biggest risk to vaccines in the cold chain was freezing and not heat, and that in one instance although one fourth of vaccines experienced freeze damage, only 5% of fridges were responsible (Hubbard 2020). Improved monitoring and data analysis of disruptions to the cold chain can greatly improve outcomes. FedEx has also launched its own sensor tracking technology, SenseAware ID, with healthcare, aerospace and retail industries expected to receive enhanced data on their shipments starting in November 2020, including those requiring cold chain, thermal blankets, and temperature controlled environments (FedEx 2020).

Some research also attempts to circumvent the cold chain in its entirety, through such means as eliminating the need for adjuvant (Sun et al. 2016) or using dried viral vaccines in  a pullulan and trehalose mixture (Leung et al. 2019). The latter enables the vaccines studied to retain efficacy for at least two months at 40 degrees Celsius and at least 3 months at 40 degrees Celsius through the use of prior approved FDA materials (Leung et al. 2019). However, despite the use of existing materials, the research team acknowledges that the pharmaceutical and logistics industry, as well as NGOs and governing bodies, are already heavily invested in the cold chain (Cooney 2019). Therefore, even given the simple technical solution, the McMaster University research group does not anticipate a large-scale vaccine supply chain shift (Cooney 2019).

*2.3 Last Mile*

Of particular importance for the vaccine supply chain, further differentiating it from traditionally studied supply chains, is the "last mile." This encompasses getting the vaccine product to the end user in a medically compliant manner. One CDC estimate puts the value of vaccines destroyed due to improper storage and transportation at $300 million per year (Gunn 2020). Similarly, one study out of Ontario, Canada put 20% of the points of vaccination (physician offices and healthcare facilities) as noncompliant with vaccine storage and handling, amounting to $3 million of wasted vaccines annually in Ontario alone (Weir and Hatch 2004). The last mile not only has monetary impacts, but severe impacts on human health. For example, in Kapoeta, South Sudan 15 children died of improperly stored measles vaccines in May 2017 (Gunn 2020). Without simultaneously understanding the network required for the vaccine supply chain and how to plan, absorb, recover, adapt to disruptions, human life is at stake.

In tandem with new technologies such as drone delivery (Forde 2019), new vaccine management strategies such as "Controlled Temperature Chain" (CTC) are also being developed to improve economic viability and outreach of vaccine supply chains, especially in the last mile (Controlled 2017). Vaccines developed and subsequently labeled for CTC supply chains are able to enter ambient temperatures less than 40°C for a number of days prior to administration, simplifying storage, transportation and time constraints of the last leg of the traditional cold chain, as well as eliminating the need for medical worker time spent conditioning ice packs prior to vaccination campaigns (Controlled 2017).

*2.4 Reverse Logistics and (biohazard) Waste Management*

Due to various CCE requirements, such as reuse of phase changing materials (PCM) in passive cold chain, and medical waste requirements of immunization equipment and associated medical devices, models of the vaccine supply chain must also specifically incorporate waste management and reverse logistics into their networks. Mass vaccine campaigns must ensure that not only the proper protocols for cold chain storage are maintained, but also proper protocols are followed for waste disposal as new vaccination centers are set up (Toner et al. 2020). Furthermore, due to the sensitive nature of vaccines, strict adherence to expiration dates must be followed and as such pharmaceutical reverse distributor programs are available to collect unused single-dose or multi-dose vials, and manufacturer-filled syringes of vaccine or diluent (MHS 2020). One development that would impact waste management is the technology for needleless flu vaccines. These are envisioned to be dispensed as patches that do not require cold chain, can be administered in rapid large batches, and do not require disposal or reuse of needles (Hayes 2019; Wedlock 2019).

Another nuance of the vaccine supply chain is the likelihood of wasted vaccine itself, or “open vial waste” and the trade-offs between single-dose formats and multi-dose formats. Such trade-offs include the higher filling costs, vaccine overfill adjustments, storage requirements, medical waste and packaging costs for lower dose vial formats, and the “open vial waste” of higher dose vials (Lee et al. 2010; Assi et al. 2011; Haidari et al. 2015). Some vaccines are packaged as multi-dose vials for use during a vaccination session intended to serve more than one individual, but if the number of doses within the vial does not match the number of people at the session, then the remaining vaccine must be disposed of and is termed “open vial waste” (Chen et al. 2015). Lee et al. use an economic computational model do understand thresholds for single-dose and multi-dose vaccine formats for measles (MEA), Hemophilus influenzae type B (HiB), Bacille Calmatte-Guerin (BCG), yellow fever (YF) and pentavalent (DTB-HepB-Hib) (Lee et al. 2010). They find that for MEA, the single-dose vial should be used for up to 2 patients per day and the 10-dose above; for BCG the 10-dose vial should be used for up to 6 patients per day and the 20-dose above; for Hib the single-dose vial should be used for up to 5 patients per day and the 10-dose above; for YF the 5-dose vial should be used for up to 33 patients per day and the 50-dose above; for DTB-HepB-Hib the single-dose vial should be used for up to 5 patients per day and the 10-dose vial above (Lee et al. 2010). It is important to note that this model does not consider the increased chance of contamination with muli-dose administration (Lee et al. 2010). Building on this, Haidari et al. address optimal primary vaccine container size by also specifically modeling the Benin vaccine supply chain through the HERMES software platform, finding that the larger dose containers reduced supply chain bottlenecks, but ultimately recommending that vaccine supply chains be individually modeled similar to other industries in accordance with specific locations, stakeholder goals, government policies, populations and health worker training (Haidari et al. 2015). Despite using a more complex supply chain model with transportation and C2 characterizations, similar to Lee et al., Haidari et al. also neglect to include the greater user error and propensity for contamination found in large-dose containers (Lee et al. 2010; Haidari et al. 2015). Although these results are intended to guide vaccine developers, manufacturers, distributors, and purchasers through larger network analyses, disruptions – anticipated or not – have not been considered, which could compromise a vaccine program (Lee et al. 2010).

*2.5 Vaccine Supply Chains and Interconnected Networks*

Although vaccine supply chains do not operate independently of other networks, such as transportation and public health policy (i.e. C2 network), few publications directly address the associated network intricacies of the vaccine supply chain. One study analyzes existing transportation and storage capacity of the vaccine supply chain in Thailand with varying time frames and target populations, finding that transportation bottlenecks are a significant issue (Assi et al. 2012). Another example is the United States health care supply chain, which relies heavily on Chinese manufacturing, which plays an interconnected role with vaccine administration for products such as PPE (Sutter et al. 2020).

In the United States, of the pharmaceutical companies that responded to the 2019 HDA Research Foundation survey, 100% of distributors stock cold chain products, with distributors ranging from traditional, specialty to third party logistics providers (HDA 2019). However, only 40% responded that they monitor and record the temperature of products in transit (HDA 2019). If manufacturers are to expand their shipments of temperature dependent vaccines in response to global epidemics, the distributors they rely on for normal pharmaceutical supply chains will need to ensure their cold chains are able to handle the demand, and necessitates incorporation into the supply chain models.

Optimization tools can also be used to coordinate supply chains relying on similar underlying networks. Due to the cold chain infrastructure required of the vaccine supply chain, the earlier comparisons with agriculture can be extended, with the food supply cold chain and global efforts to minimize food waste and capacitate rural farmers (UN 2020). Initiatives such as the United Nations Environmental Program are combining efforts in their cold chain initiatives so that food and vaccine can use the same supply chains, calling for “resilient, reliable and sustainable cold chains” and seek to pivot away from traditional methods of large-scale refrigeration in order to reduce fossil fuel and refrigerant use, while increasing availability (UN 2020). Analytic tools including value of information on vaccine inventory levels, and trade-offs with visibility and cost are also examples that can be used (Li et al. 2018).

Public opinion towards vaccines is also an associated network that has had impacts on successful vaccination campaigns, with misinformation expected to increase as new vaccines are developed (Larsson 2020). For example, the New Jersey Department of Health found that once available, only 60% of surveyed physicians and 40% of surveyed nurses would get vaccinated (Walsh 2020).

*2.6 MMR And Smallpox Vaccine Case Studies*

The measles-mumps-rubella and smallpox vaccination programs underscore the importance of the entire vaccine supply chain and the associated networks and domains such as transportation, social, C2, etc. Measles was responsible for over two million deaths annually prior to the introduction of the Expanded Program on Immunization (EPI) in the 1980s (WHO 2019). Although current vaccine immunization programs and supplementary mass preventive vaccination campaigns are estimated to have prevented 21.1 million deaths globally, annual deaths due to measles remain at 100,000 (WHO 2019). Despite an efficacious vaccine, rubella also continues to be a global health priority (WHO 2019). Due to high transmission rates, measles and rubella outbreaks are used by public health officials to assess overall immunization gaps (WHO 2019). This ultimately serves as vaccine supply chain viability litmus tests, which could also potentially be linked to disruption.

As of September 2019, 82 countries were considered to have eradicated measles and 81 rubella (WHO 2019). Country eradication can be reversed, however, as countries face insufficient political will, conflict, migration, humanitarian emergencies, lack of vaccine investment and “vaccine hesitancy” (WHO 2019). The M-M-R II vaccine requires cold chain, but does not lose potency if frozen, making it less susceptible to disruptions along the supply chain (DHA 2019). In fact, it can be used to buffer other vaccines from freezing (WHO 2015). Seven countries once having been declared measles free have already re-established virus transmission, regardless of this flexibility in the cold chain (WHO 2019). As such, an MMR vaccine that is independent of the cold chain or skilled health-care workers for distribution would be advantageous on the global scale (Lambert et al. 2015).

Smallpox, on the other hand, was considered completely eradicated when the 33rd World Health Assembly, on May 8, 1980, officially declared the world free of smallpox (CDC 2016). Smallpox is transmissible from human to human in dense populations, but is not considered to be highly transmissible and outbreaks were not associated with locations like schools or trains; though three out of ten contracting the disease died (Belongia and Naleway 2003; Peterson et al. 2015; CDC 2016). Although mass vaccination was the original approach, the smallpox eradication program ultimately employed the "ring" vaccination strategy to reach complete eradication, strategically targeting hot spots and contacts of known cases to break the chain of transmission (Belongia and Naleway 2003; CDC 2016; Toner et al. 2020). The eradication program was initiated by the World Health Organization in 1959, but a lack of funds and personnel rendered it unsuccessful until it was relaunched in 1967 as the Intensified Eradication Program with the goal of 80% vaccine coverage in every country through increased global laboratory coverage providing higher-quality vaccine, development of the bifurcated needle, and use of the ring vaccination or surveillance system (Belongia and Naleway 2003; CDC 2016).

Dryvax, the vaccine that was used for the smallpox eradication programs, does require refrigeration, and cannot be frozen, but the success of the eradication campaign is partially attributed to its long-term stability outside of cold temperatures (Belongia and Naleway 2003; Wyeth 2004). More recently, Dryvax was replaced by a couple other vaccines with fewer side effects, including ACAM2000 in the U.S. strategic stockpile in case of bioterrorism, but continues to require the cold chain (Peterson et al. 2015; Emergent 2018). The global eradication of smallpox is considered by some one of the greatest achievements in human history (Belongia and Naleway 2003). Not only the manufacturing and technological advances in vaccines were vital, but smart distribution and last mile policies and supply chains.

### *2.7 Vaccine Supply Chain Disruptions*

Most disruptions in the vaccine supply chain have been studied in the context of expected disruption to power supply in the cold chain. Chen et al. develop a model addressing cost tradeoffs for passive CCE design within the Niger vaccine supply chain with variables including ice recharging, vaccine storage space, storage volume, doses per vial of vaccine, vial volume, truck loads, vaccine packing ratio, and ice packing ratio, among other variables (Chen et al. 2015). There are historic examples of short-term disruptions, however. In 2008, Turkey experienced a disruption to its vaccine supply chain when the first generation of pneumococcal vaccine was introduced, more than quadrupling their cold storage requirements from 2,600 square meters to 11,400 square meters, and ultimately had to rent cold storage space (Humphreys 2011). And in 2016, an explosion at a Chinese factory caused a shortage of the broad-spectrum antibiotic piperacillin-tazobactam due to traditional manufacturing of active pharmaceutical ingredients (APIs) localized at only a few factories (Cogan et al. 2018). The 2017 back-to-back hurricanes Irma and Maria devastated Puerto Rico, which in 2017 accommodated over 70 medical device manufacturers and 49 pharmaceutical companies, producing about 10% of the U.S. drug supply and a larger share of the U.S. IV solutions (Stone 2018). In response to the subsequent shortage, the FDA approved imports from additional countries and animal sources, while also extending the expiration date on available IV fluids (Hayes 2018; Stone 2018).

The vaccine supply chain is also especially susceptible to manufacturing disruptions as specific vaccines rely on specific processes and therefore specific and sustained raw materials, meaning that competitive pressure from within the industry or even other industries could not only potentially increase cost, but also interrupt the supply chain (Plotkin et al. 2017). Global vaccine availability hinges on the delivery of potent and effective vaccines at point of use through reliable distribution channels, strict quality control and in-depth production methods and raw material sourcing at manufacture (Smith et al. 2011). Consequently, a disruption at any node or link could have devastating human impacts. Even so, most of the existing literature looks at optimizing the reach of vaccines in developing countries under known and expected disruptions (e.g. unreliable electricity grids and intermittent transportation infrastructure), but what about unanticipated disruptions to the global vaccine supply chain and their associated networks? Disruptions are inevitable in supply chains of this magnitude, underscoring the need for ensuring resiliency of the network (Linkov 2020). These "unknown unknowns" can only be understood through resilience analytics of vaccine supply chain models.

Analyzing and modeling how a supply chain will react to disruption is vital to ensuring distribution targets are met. A resilient vaccine supply chain will continue to achieve immunization targets post disruption while a less resilient supply chain may leave already marginalized populations without recommended immunizations (see *Figure 2* for depiction).

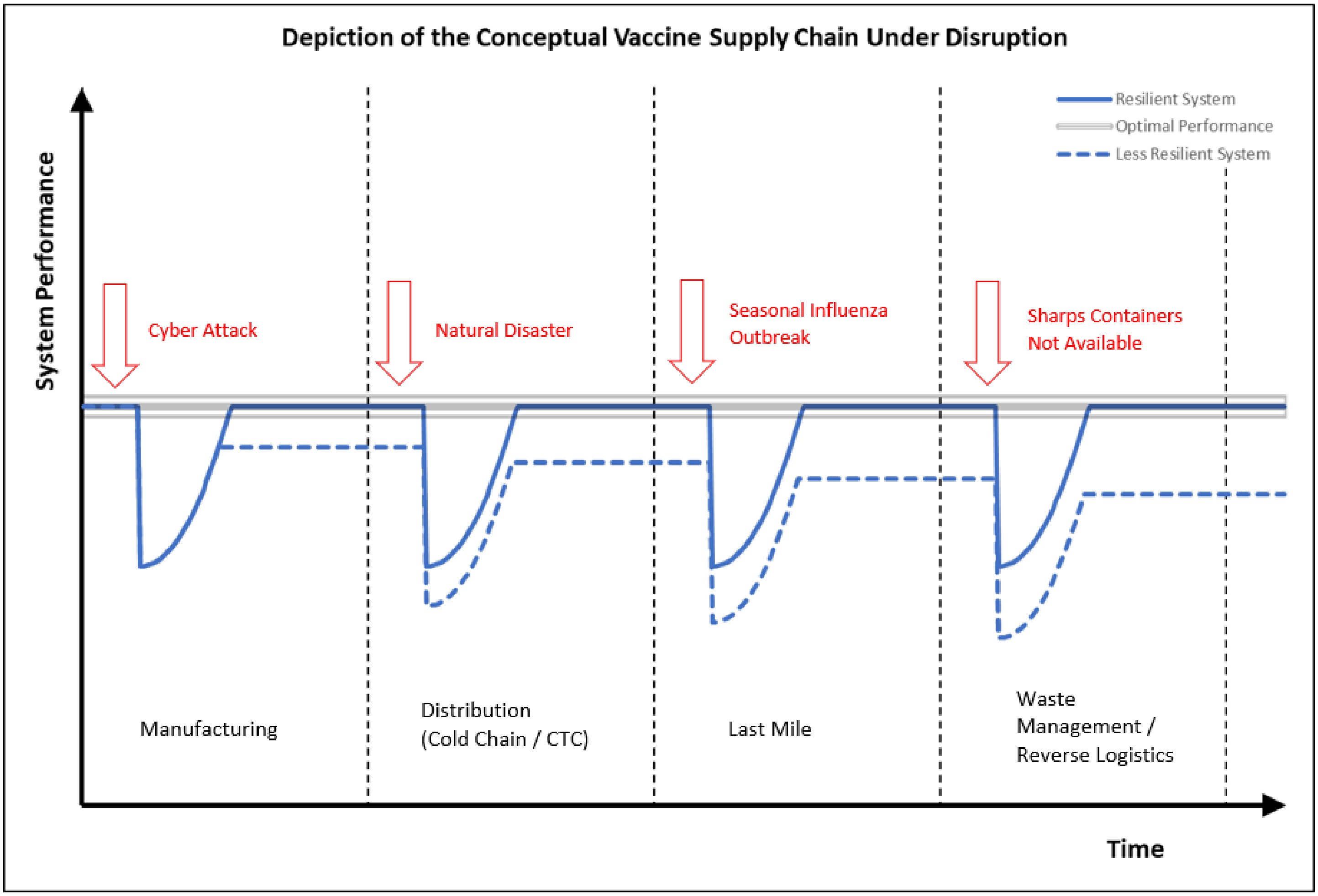

*Figure 2: Notional comparison of a resilient and less resilient vaccine supply chain experiencing examples of disruptions during different stages of the vaccine supply network. Note these possible disruptions are not unique to any stage.*

## III. Supply Chain Models and Resilience Analytics

An efficient and resilient vaccine supply chain is fundamental for achieving a wide-spread and equitable immunization response. Although at the heart of any immunization program is the vaccine itself, experts in the immunization field have the mantra, '*no product, no program'* calling attention to the fact that the network of staff, equipment, vehicles, and data is equally as important as the existence of a vaccine (Moeti et al. 2017). In other words, without a supply chain and its associated networks, a vaccine cannot meet intended public health goals. And even with an efficient supply chain, if it is not resilient, it will not maintain optimal functionality after an inevitable disruption. As such, quantitative vaccine chain models and resilience analytics are warranted.

*3.1 Lack of Existing Academic Publications on Vaccine Supply Chain Resilience*

A topic search in Web of Science (WOS) "All Databases" for relevance to "supply chain*", "vaccin*" and "resilien*" returned only two results. The first publication is a review article that does not provide a fundamental definition of resilience and minimally addresses supply chain, but does acknowledge the constraints of cold chain (Tambo et al. 2018). The authors do point out "strengthening acceptable and effective intervention packages requires multidisciplinary and intersectoral linkage," but offer no models or quantitative methods for policy makers (Tambo et al. 2018). The second publication focuses on veterinary vaccine supply chains and does not offer metrics for measuring supply chain resilience (Dungu 2020). However, Dungu does offer a valuable discussion on types of vaccine banks that improve resilience of vaccine supply chains, by defining (1) physical vaccine banks – storage of ready-to-use single or multi dose vaccines; (2) virtual vaccine banks – an agreed upon amount of vaccine can be produced and ready for distribution in a certain amount of time should the need arise; (3) maintenance of production capacity – consistent management of vaccine seed material and ensuring supply chains are in place (2020). The latter is most critical for pandemics as existing infrastructure and material could be adapted to any disease. For example, the U.S. keeps a stockpile of chicken eggs for use for influenza vaccines (BARDA 2020). However, this hedge will not help for newer vaccine development technologies, such as many of the coronavirus vaccines under development, as different vaccine mechanisms are being used (Yeung 2020). New vaccine technologies, such as those that are platform-based, offer rapid scale and delivery, which could be vital for stockpile replenishment during outbreaks, offering a "warm-base" (Adalja et al. 2019), potentially hardening the vaccine manufacturing ecosystem against risk and increasing the resilience of the vaccine supply chain.

Due to the lack of academic publications, a media search using the same WOS topic search criteria was conducted on October 19, 2020 using the Google News search engine for historical searches for each month from October 2010 to September 2020. Media attention specifically pertaining to the vaccine supply chain and resilience has been on the rise (see *Figure 3*), with many news sources focused on humanitarian vaccination and immunization developments, especially with the rise of vaccine initiatives by the Bill and Melinda Gates Foundation and its partners. Some of these publications look at emerging technologies in vaccine development, while others promote the benefits of private-public cooperation (Hoybraten 2014). The Ebola epidemic is also a common theme. There is also a theme among news feeds showcasing drawbacks of concentrated manufacture in China, including the presence of contaminated vaccines and further need for oversight at the beginning of the supply chain (SDC 2018). Since the beginning of the COVID-19 pandemic, google news searches have returned articles pertaining to expected logistics shortcomings and supply chain bottlenecks for implementing a vaccination campaign once the vaccine has been approved. Of note, some U.S. publications did point out shortcomings in the existing domestic supply chain prior to the pandemic, such as "Fragile Antibiotic supply Chain Causes Shortages and Is a National Security Threat," which points to the fact that 80% of the U.S. pharmaceutical raw materials come from China and India (Stone 2018). One estimate puts the number of drug shortages as having tripled between 2006 and 2018, putting blame on pharmaceutical manufacturers moving to countries such as China and Brazil, citing issues such as security, intellectual property and conflicting government policies (Morris and Sweeney 2019). One article does analyze uncertainty in the vaccine supply chain (Comes et al. 2018). Although Comes et al. do not specifically look at resilience in vaccine supply chains, they show the need for planning and implementing cold chains under uncertainty, through two complementary approaches: "adaptive policymaking" and "adaptation pathways" (Comes et al. 2018).

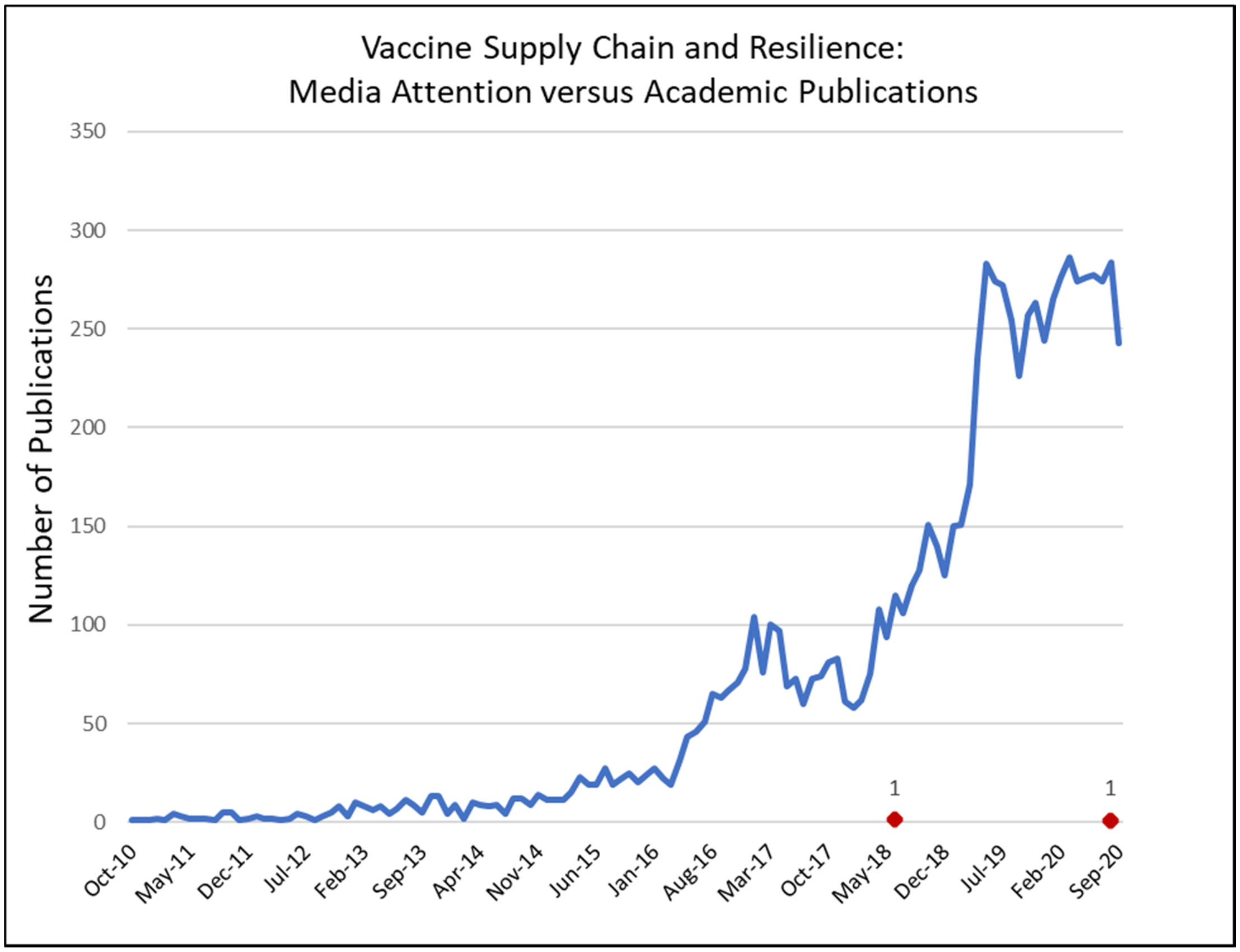

*Figure 3: Number of available Google News results on a monthly basis (blue) as compared to published scholarly articles found on Web of Science (red), showing lag in academic research.*

Resilience of the vaccine supply chain is critical. However, a lack of models, metrics and network analytics continues to plague the academic vaccine supply chain resiliency field.

*3.2 Supply Chain Resilience Publications – Modeling Trends*

Although the vaccine supply chain presents unique challenges, prior research on quantitatively modeling resilience in supply chains shows (1) a lack of the four-stage NAS definition of resilience (plan, absorb, recover, adapt); (2) a lack of modeling disruptions of different magnitude, likelihood and systemic threats; (3) lack of the tiered approach to modeling; (4) lack of modeling associated networks that constitute value generation (e.g. C2 – command and control, transportation, cyber) (Golan et al. 2020; Mersky et al. 2020).

The supply chain resilience field is rapidly growing, as indicated by the Web of Science "All Databases" search, with the share of supply chain publications over the past ten years discussing resilience jumping from 0.7% in 2010 to 4.2% in 2020 (*Figure 4a*). Similarly, the portion of resilience publications focused on supply chains have also increased from 0.3% in 2010 to 1.3% in 2020 (*Figure 4b*). Note that these publications may not use the term "resilience" as defined by NAS, but have been tagged as relevant by Web of Science, indicating relative trends in the field.

Of note, although not included in the academic search, there are some private companies that offer resilience benchmarks to their clients. Resilinc Corporation, for example, emphasizes its capability to map multiple tiers of the supply chain and monitor disruptions through social media monitoring (Resilinc 2017). This emphasis on visibility allows Resilinc's clients in the pharmaceutical industry to quickly pivot to other sources as soon as a possible disruption is detected in a supplier they may not have otherwise known is affiliated with their supply chain, such as with Biogen during hurricanes Harvey, Maria and Irma hitting Puerto Rico (Resilinc 2017). This R Score™ can range on a scale of 1 to 10 and is a weighted average that factors (1) transparency of information; (2) network locations in relation to factors such as geographical dispersion, geopolitical stability, natural disaster resistance, and macro-economic strength; (3) continuity of recovery time capabilities at company sites; (4) performance regarding financial stability, quality, and responsiveness; (5) supply chain resiliency program maturity (SCRM), which encompasses ongoing efforts by the company for supply chain visibility and monitoring, and proactive risk management (Resilinc 2017). This score can then be used to compare improvements quarterly within a company as well as benchmark against other companies within the same industry to improve supply chain resilience.

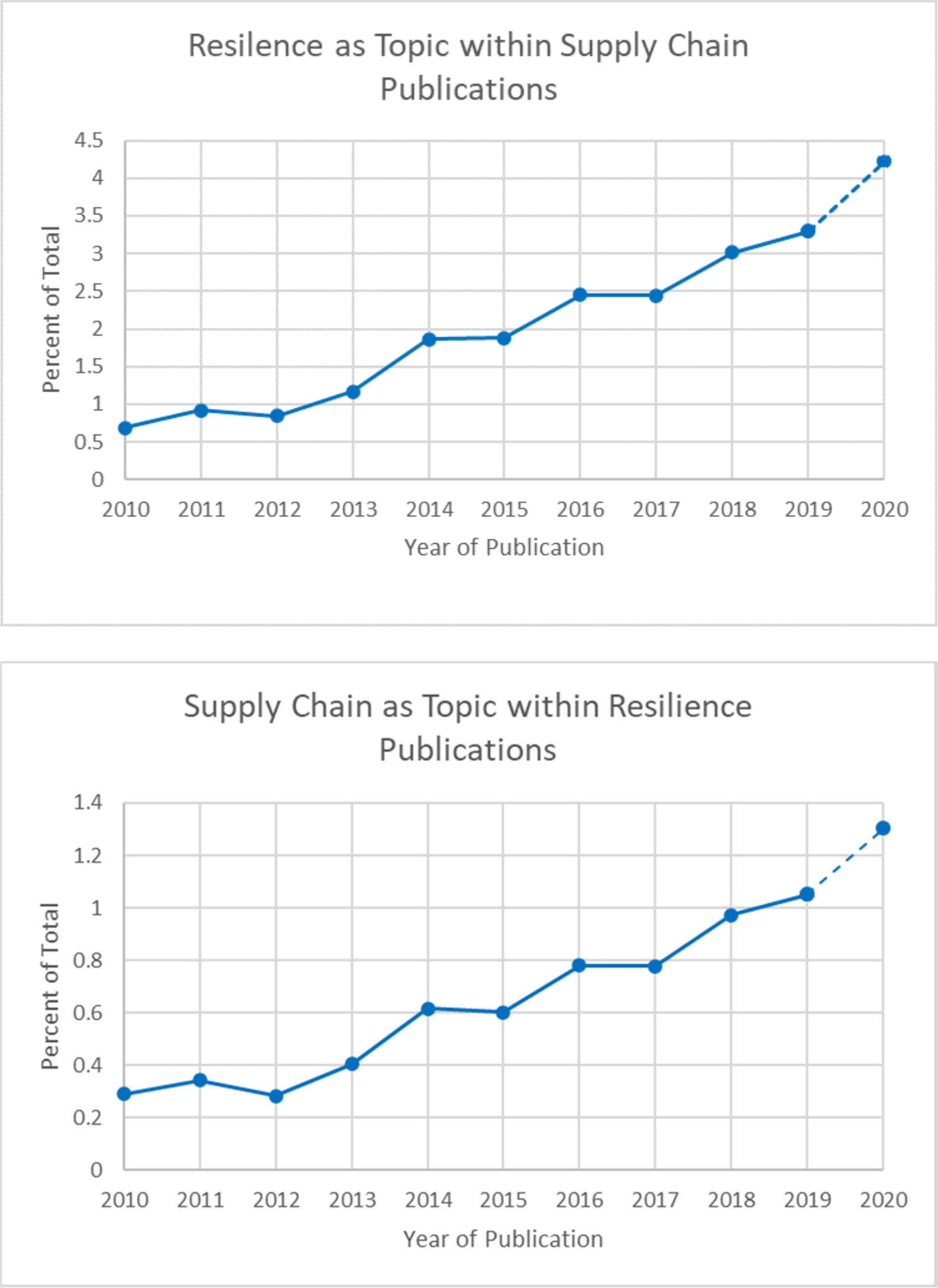

*Figure 4: Share of Supply Chain publications discussing Resilience (a) and share of Resilience Publications Discussing Supply Chains (b) from WOS October 11, 2020 topic searches. Data for 2020 may be incomplete.*

*3.3 COVID-19 Current Trends and Impacts on Supply Chain Resilience*

Understanding the system dynamics from disruptions due to the COVID-19 pandemic on existing supply chains is essential to developing resilient vaccine supply chains that must achieve optimal performance while experiencing impacts from the pandemic itself. Building on the results from the 2020 review of resilience analytics in supply chain modeling for trends through the end of 2019 (see Golan et.al., 2020 for full discussion), publications from the year 2020 were analyzed with the additional cross section of the COVID-19 Pandemic. Specifically, the search for 2020 was conducted in Web of Science "All Databases" during the second week of October 2020. All English language publications found under the Topic Search = ("supply chain" AND "resilien*" AND "covid*") were filtered for a minimum of 2 citations. This resulted in 12 publications for review. To check that no articles were missed, the Topic Search = ("supply chain" AND "resilien*" AND "pandemic*") was also run. This resulted in an additional article published in 2018, which did not meet citation requirements. Although the number of papers reviewed for 2020 may be underestimated, it captures the growth in publications pertaining to supply chain resilience, especially during initial analysis of the impacts of the COVID-19 pandemic on supply chain resilience. Of the 271 publications from 2020 tagged under supply chain resilience in WOS, 42 are also tagged with "covid*" (15%) and 35 are also tagged with "pandemic*" (13%). However, despite the burgeoning literature, only 12 of these publications met citation requirements and only 4 met relevancy requirements for review.

The four additional publications analyzing supply chain resilience in the context of COVID-19 extend the trends identified in earlier literature, and further highlight the gaps in current understanding and models of supply chain resilience. None of the publications take a direct approach to defining resilience, missing an opportunity for clear 4-stage temporal analysis mid-pandemic. Although all relevant publications discuss at least two stages of resilience, there is overwhelming focus on the absorb and recover stage (see *Table 4*).

*Table 4: Resilience Characteristic Results for 2020*

| Publication | Plan | Absorb | Recover | Adapt | Metric | SC | Trans. | Decision | Scenario |
|---|---|---|---|---|---|---|---|---|---|
| Ivanov, D. March 2020 (a) | No | Yes | Yes | No | Proxy | Graph | Graph | Optimization | Case study |
| Ivanov, D. May 2020 (b) | No | Yes | Yes | Yes | Proxy | Graph | Graph | Optimization | Set list |
| Ivanov, D. & Dolgui, A. April 2020 (a) | No | Yes | Yes | Yes | No | Graph | None | Optimization | None |
| Ivanov, D. & Dolgui, A. May 2020 (b) | No | Yes | Yes | No | No | Graph | Graph | Optimization | Case study |

Although the publications span multiple sectors and locations, the analyses lack clear insights into modeling resilient supply chains and miss the opportunity to provide metrics in real time. Of the two publications that use case studies, one is unrelated to a pandemic disruption (Ivanov and Dolgui 2020b), and one analyzes lightning equipment in China by proposing comparisons of performance quantifications without a direct connection to supply chain resilience, but rather indirectly through inventory, customer, financial and lead-time performance levels (Ivanov 2020a). Another publication uses set lists, but minimally analyzes hypothetical disruptions, while also measuring supply chain performance indicators such as cost and service level (Ivanov 2020b). The author neither tests the model in the context of a real-world disaster, nor offers a true model for supply chain resilience analytics. However, the focus on “Intertwined Supply Networks” does add depth to the supply chain model, and shows that unanticipated connections among healthcare, industrial, pharmaceutical and food supply chains are necessary to include in supply chain models. Overall, the clear lack of resilience metrics for supply chains apparent in the 2007-2019 literature review hold true despite the plethora of various disruption types caused by the COVID-19 pandemic.

## IV. COVID-19 Vaccine Supply Chain Resilience Needs and Challenges

Some estimates indicate that 70% of the global population will need a COVID-19 vaccine in a short timeframe, which will overwhelm the existing infrastructure and vaccine supply chain systems in place (UN 2020). Assuming that manufacturing capacity is scaled, should a disruption in the cold chain occur, for example, spoilage on the order of billions could occur if current global vaccine spoilage rates hold true (UN 2020). Even prior to approval of a vaccine and Phase IV of the vaccine development process, clinical trials are also reliant on a working vaccine supply chain. It is therefore imperative to proactively model the COVID-19 vaccine supply chain with resilience analytics so that vaccine manufacture, distribution and inoculation can occur regardless of disruption events.

Operation Warp Speed in the U.S., for example, addresses vaccine development, manufacture and distribution, with the goal of 300 million vaccine doses delivered by January 2021 (HHS 2020b). Such initiatives include the preemptive manufacture of needles, syringes, vials, supply kits and fill-finish equipment (HHS 2020c). These consumables are being produced under Federal contract for the Strategic National Stockpile to have 400M to 700M each by the end of 2020 (Elton et al. 2020). For example, SiO2 – one of two leading glass vial manufacturers in the U.S., along with Corning – has increased production to 120M vials per month (Elton et al. 2020). The overview of distribution and administration for Operation Warp Speed shows a linear flow of material from Contracted OWS Manufacturers to Distributors to Partner Depots and finally to Administration Sites (Pharmacy, LTC Providers, Home Health, Indian Health Services, Other

Federal Entity Sites, Public Health Clinics, Hospitals, Doctor's Office, Mobile Vaccination, Mass Vaccination), including a side distribution for Ancillary Supplies/PPE to Kitting and Distributor (HHS 2020a).

Therefore, similar to other national and global efforts (e.g. COVAX), Operation Warp Speed is directly supporting vaccine efforts (HHS 2020b). The new platform vaccine technologies being developed – emerging infectious disease (EID) medical countermeasures (MCMs) – do provide economies of scale, but by the very nature of EIDs, will not provide financial rewards without private-public partnerships (Adalja et al. 2020). *Table 5* highlights the supply chain ecosystems supported by the U.S. Government through partnerships with manufacturers for accelerated vaccine production. Although they incorporate platform technologies which were already in place and available for quick production, they are not commercially optimized, all requiring cold chain and multiple doses (Gottlieb 2020). The number of doses implicates not only the distribution mechanisms of the supply chain, including cold chain, last mile, and associated networks, but the supply chains underpinning the manufacturing ecosystem, such as syringes, vials, and biologicals.

*Table 5. Overview of vaccines supported by OWS or other U.S. federal government efforts (CoVPN 2020; Folegatti 2020; Jackson et al. 2020; Keech et al. 2020; Sadoff, et al. 2020; Sahin et al. 2020; WHO 2020)*

| Manufacturer | Vaccine Name(s) | Vaccine Type | Number of Doses | Notes |
|---|---|---|---|---|
| AstraZeneca / University of Oxford | AZD1222<br>ChAdOx1-S | Recombinant vector (platform based) | 2 (day 1 and 29) | • Uses chimpanzee adenovirus to deliver SARS-CoV-2 spike protein<br>• ChAdOx1 immunogenic in older adults (i.e. target population) and can be manufactured at scale |
| Janssen / Johnson & Johnson | Ad26.COV2.S<br>JNJ-78436735<br>Ad26COVS1 | Non-replicating viral vector (platform based) | 1 for lessening severity<br>2 (day 1 and 56) | • ENSEMBLE study<br>• Uses Janssen's adenovirus vector to deliver SARS-CoV-2 spike protein<br>• Promising immunogenicity and manufacturing profiles |
| Moderna | mRNA-1273 | LNP-encapsulated mRNA (platform based) | 2 (day 1 and 29) | • COVE study<br>• mRNA encodes the SARS-CoV-2 spike protein |
| Novavax | NVX-CoV2373 | Full length recombinant (platform based) | 2 (day 1 and 22) | • SARS-CoV-2 spike protein in baculovirus expression system<br>• Matrix-M1 adjuvant (saponin based) for thermostability (2 to 8 degrees C) |
| Pfizer & BioNTech | BNT162b2<br>BNT162b1 | 3 LNP-mRNAs (platform based) | 2 (dose finding) | • Lipid nanoparticle-formulated mRNA with SARS-CoV-2 spike protein<br>• Quick and scalable mRNA manufacturing & LNP formulation |
| Sanofi / GlaxoSmithKline | SARS-CoV-2 biological & adjuvant formulations | Recombinant Protein (platform based) | 2 (day 1 and 22) | • SARS-CoV-2 spike protein in baculovirus expression system<br>• Uses GSK established adjuvant |

As implementation of Operation Warp Speed continues with increased manufacturing capacity, attention to resiliency against disruptions is necessary, not just at specific manufacturing nodes, but throughout the entirety of the associated networks, such as transportation and C2 (Golan et.al. 2020). This includes anticipating how other essential vaccine supply chains will be affected, as goals for one vaccine campaign may impact all vaccine supply chains (Assi et al. 2012). And conversely, there is also the opportunity to capitalize on the large COVID-19 vaccine campaign and administer other necessary immunizations simultaneously (Toner et al. 2020). Other critical infrastructure functions must also be maintained, and security risks minimized (Gomez et al. 2020). During manufacture, some risks for OWS include cyber-attacks and biological threats. For example, as of October 8, 2020, several U.S. vaccine manufacturers and academic labs have been targets by Chinese government-linked hackers (Sutter et al. 2020).

As discussed in prior sections, the last leg of the vaccine supply chain will not only be controlled by the associated networks, but specifically largely controlled by policy makers (C2), who will implement allocation to targeted populations as well as control vaccine manufacture and supply. Beyond specific populations deemed essential for national security (e.g. military, etc.), understanding which populations are at the highest risk is also essential to an inoculation program and targeted supply chain – adding value to the "value chain" (Linkov et al. 2020). Populations at higher risk include those with more severe symptoms due to physical underlying conditions (e.g. age, weight, respiratory), but also from socioeconomic and demographic trends, including homeless populations. In the U.S., for example, the country leading the globe for both the number of confirmed COVID-19 cases and deaths, Latinx people are hospitalized at 4 times the rate of their white counterparts (Watson et al. 2020). The trend is even higher in Black, Native American, and Alaska Native people who are hospitalized at 5 times the rate of their white counterparts (Watson et al. 2020). Additionally, the transmission rate among incarcerated individuals is more than double that of the U.S. population, and within Immigration and Customs Enforcement detention centers, models show infection of 72% to 100% of individuals within 3 months of the first infection (Watson et al. 2020). This has larger implications on the population as a whole and associated networks. For example, as of April 19, 2020, up to 15.9% of all COVID-19 cases in Chicago may be attributed to Chicago's Cook County Jail (Watson et al. 2020).

Another Johns Hopkins study addresses priority groups for the vaccine response, incorporating the unequal morbidity and mortality rates across population sectors into their model, suggesting that Tier 1 include those groups (1) essential in sustaining the ongoing COVID-19 response, (2) at greatest risk of severe illness and death, and their caregivers, (3) most essential to maintaining core societal functions; and Tier 2 include those groups (1) essential to broader health provision, (2) with least access to health care, (3) needed to maintain other essential services, (4) elevated risk of infection (Toner et al. 2020). Understanding the intricacies of the last mile for targeted populations is necessary to incorporate in the vaccine supply chain model, allowing for effective targeted vaccination campaigns such as the smallpox “ring” strategy. Targeting these priority populations requires an overarching national strategy, that could potentially allocate vaccines to states on a per capita basis or hot spot basis (Toner et al. 2020). The supply chain needs to be able to respond to any ebbs and flows in disease outbreaks and/or changes in distribution policies (agile and resilient).

Also at play in public policy and is the notion of “essential.” Different definitions of “essential workers” throughout individual state responses may muddle a national response once a vaccine is ready for distribution. Twenty-two of the 42 states with essential worker orders differ from the federal definition given by the US Cybersecurity and Infrastructure Security Agency (CISA) (NCSL 2020). Proper public education campaigns should be coordinated to minimize last mile disruptions. Finally, there will need to be a way to track who has received a vaccine, which will be especially critical if efficacy requires two doses (Toner et al. 2020). As with any health tracking policies, health privacy issues would need to be addressed. Impacts from concurrent public health policy objectives such as social distancing must also be accounted for in vaccine supply chain models, as social distancing requirements, for example, could decrease the number of people able to queue for vaccine distribution and the amount of time and economic input required by health professionals for sterilization between patients.

Understanding who the end user is, and how to mitigate disruptions in the last mile of delivery to high-risk populations is imperative in proactively modeling and developing a resilient vaccine value chain. Even if an efficient supply chain is developed, if a disruption leaves it below optimal performance, widespread and timely distribution of the vaccine may not be achieved: a resilient vaccine supply chain model is warranted (Linkov 2020).

## V. Conclusions

In light of the current COVID-19 pandemic and the rush to bring a viable vaccine to market, ensuring an efficient mechanism is also in place for distribution to target populations in an equitable and resilient manner can only be achieved with resilience analytics in vaccine supply chain network models. Unexpected disruptions in such a large-scale effort are inevitable; using the four phase temporal approach to resilience – plan, absorb, recover, adapt – to quantitatively model all facets of the vaccine supply chain and its associated networks is missing in the current vaccine supply chain literature and needs to occur as aggressively as vaccine development. All supply (value) chains operate within system domains: physical, cyber, cognitive and social (Golan et al. 2020; Linkov et al. 2020; Mersky et al. 2020) and this must be expanded to vaccine/immunization supply chains in order to fully understand the network interactions and ultimately vaccine supply chain resilience in a quantitative manner.

As emergency use authorization (EUA) by the FDA for a COVID-19 vaccine is likely (Johns Hopkins 2020), the vaccine supply chain will also need to be efficient enough to ramp up production and logistics, while also able to be resilient in the face of likely disruptions, ensuring target populations are inoculated in a timely manner and lives saved. A national strategy that maximizes existing networks of both existing vaccine supply chains and the associated networks will be necessary. Consideration of the consequences of unequal disruptions and cascading failures on marginalized sectors of supply chains is necessary in order to support resiliency of the entire supply chain network and ensure that goals of the networks are met.

Given recent promising vaccine technologies, from the form the vaccine takes, such as needleless, dried, or ambient temperature resistant, to the types of cold chain equipment used in distribution and the last mile, it is more imperative than ever to model the larger network interactions and quantify where to invest time, energy and money to get the most value out of the COVID-19 vaccine supply chain. The supply chains underpinning the unique biopharmaceutical manufacture of vaccines to the networks ensuring safe distribution to appropriate end users must be fully modeled so that disruptions can be better understood, and the vaccine supply chain hardened in an efficient manner. Similar conclusions regarding resilience of other critical infrastructure supply chains, such as the Information and Communications Technology (ICT) and biopharmaceutical finished goods, during the COVID-19 pandemic have been reached, calling for refining supply chain risk-management approaches, mapping and standardizing detailed supply chain networks (i.e. increasing visibility into all tiers), planning for transportation bottlenecks, dual-sourcing, and holding buffer inventories (CISA 2020; Jacoby et al. 2020; Resilinc 2020).

Further building on current trends in supply chain resilience analytics modeling and prior work in the field (Golan et al. 2020; Hynes et al. 2020b; Ivanov and Dolgui 2020a; Linkov 2020; Mersky et al 2020; Trump and Linkov 2020; Linkov et al. 2018) in the context of the vaccine supply chain and disruptions discussed in this chapter, we recommend the following in the development of the supply chain underpinning the COVID-19 vaccine immunization program:

1. Incorporation of the definition of supply chain resilience across all vaccine supply chain models and sectors is necessary to make resilience management more efficient. We recommend adoption of the standard four-stage definition of resilience provided by NAS – plan, absorb, recover, adapt;

2. Consideration of different types of disruptions within the vaccine supply chain resilience models – especially assessing system recovery from unknown disruptions and systemic threats – is necessary to expand the scope that supply chain resilience management is able to quantify;
3. Consideration of the tiered approach to modeling, ranging from simple metrics to advanced network models, is necessary for understanding which quantification method to apply to the analytic need;
4. Consideration of the vaccine supply chain within the broader context of other networks that constitute value generation (e.g., command and control, cyber, transportation) and overall increased visibility and mapping is necessary for quantification of global network interactions and more robust vaccine supply chain resilience models that accurately portray trade-offs between efficiency and resilience to avoid cascading failures and maintain existing immunization goals while also meeting new public health targets;
5. Consideration of the consequences (i.e. trade-offs) of unequal disruptions and cascading failures on marginalized sectors of supply chains is necessary in order to support resiliency of the entire vaccine supply chain network and ensure that goals of the networks are met, especially in regards to critical infrastructure.

These considerations would enable public health officials, governing agencies, pharmaceutical manufacturers, and distributors, among others, to more efficiently and effectively implement COVID-19 vaccination targets and quantitatively weigh trade-offs in supply chains. Implications of the pandemic on associated networks must be modeled in tandem with the vaccine supply chain itself using resilience analytics to ensure immunization targets are met regardless of disruptions to the supply chain.